\documentclass[aps,prb,reprint,showkeys,showpacs]{revtex4-1}

\usepackage{natbib}
\usepackage{bm}
\usepackage{graphicx}
\usepackage{todonotes}
\usepackage{amsmath}
\usepackage{amsfonts}
\usepackage{xargs} 
\usepackage{color}

\usepackage{epstopdf}
\usepackage[
pdfauthor={derajan},
pdftitle={How to do this},
pdfstartview=XYZ,
bookmarks=true,
colorlinks=true,
linkcolor=blue,
urlcolor=blue,
citecolor=blue,
pdftex,
bookmarks=true,
linktocpage=true, 
hyperindex=true
]{hyperref}

\renewcommand{\r}[1]{{\color{black} #1}}
\renewcommand{\b}[1]{{\color{black} #1}}

\renewcommand{\v}[1]{{\color{black}  #1}}
\renewcommand{\gg}[1]{{\color{black}  #1}}

\begin{document}
\title{Influence of Joule heating on current-induced domain wall depinning}
\author{Simone~Moretti}
\email[Corresponding author: ]{simone.moretti@usal.es}
\author{Victor~Raposo}
\author{Eduardo~Martinez}
\affiliation{Universidad de Salamanca. Plaza de los Caidos, 37008, Salamanca. Spain.}
\date{\today}

\begin{abstract}
The domain wall depinning from a notch in a Permalloy nanostrip on top of a ${\rm SiO_2/Si}$ substrate is studied theoretically under application of static magnetic fields and the injection of short current pulses. The influence of Joule heating on current-induced domain wall depinning is explored self-consistently by coupling the magnetization dynamics in the ferromagnetic strip to the heat transport throughout the system. Our results indicate that Joule heating plays a remarkable role in these processes, resulting in a reduction in the critical depinning field and/or in a temporary destruction of the ferromagnetic order for typically injected current pulses. In agreement with experimental observations, similar pinning-depinning phase diagrams can be deduced for both current polarities when the Joule heating is taken into account. These observations, which are incompatible with the sole contribution of spin transfer torques, provide a deeper understanding of the physics underlying these processes and establish the real scope of the spin transfer torque. They are also relevant for technological applications based on current-induced domain-wall motion along soft strips.  
\end{abstract}


\maketitle

\section{Introduction}
\label{sec:introduction}

\indent Magnetic domain walls (DWs) in patterned nanostrips have attracted much attention due to their application in field- and current-induced DW logic\cite{Allwood2005} and memory devices~\cite{Parkin2008}. Key to the successful operation of these memory devices is the controllable motion of DWs between pinning sites using current pulses, which contrary to the field-driven case, will coherently drive neighboring walls in the same direction through the nanostrip. Experiments have shown that the injection of an electrical current through a soft Permalloy (Py) strip can drive domain walls in the direction of the electron flow \cite{Yamaguchi2004, Beach2006, Hayashi2007} and/or assist the field-driven depinning from a patterned constriction\cite{Hayashi2006a}. DW transformations\cite{Klaui2005}, and even the nucleation of multiple walls \cite{Yamaguchi2005} have also been observed under injection of current pulses. These observations are usually interpreted in terms of the spin-transfer torque (STT) mechanism, as predicted by Berger and Slonczewski\cite{Berger1984,Slonczewski1996}. Since the electrical current becomes spin-polarized in the magnetic direction of the ferromagnet, the polarization of the spin flips when a DW is encountered. Because of angular momentum conservation, this change in angular momentum of the conduction electrons leads to a reaction torque on the DW called adiabatic STT, and the DW is pushed in the direction of the electron flow. In addition to this torque, a second, non adiabatic STT was suggested\cite{Tatara2004,Zhang2004,Thiaville2005,Tatara2008} to explain the discrepancies with the experimental results. The common interpretation is that the spin polarization of the electric current cannot follow the local magnetization within the DW, and a misalignment between the current and the DW magnetization occurs, which acts as an additional field-like torque on the DW. This torque is parametrized by the so-called non adiabatic parameter $\xi$. However, the physical origin, as well as the magnitude of this non adiabatic STT contribution, remains controversial to this day\cite{Boulle2011}. The debate is reinforced by the high disparity of the estimated STT parameters, spin polarization $P$ and nonadiabaticity $\xi$, as deduced from experiments on similar Py samples (see \cite{Boulle2011} and references therein). Therefore, a complete understanding of these observations is lacking, and further theoretical and numerical efforts are still needed.

\indent A typical design for measuring the STT parameters consists of analysing the depinning of a DW initially trapped at a notch. In the absence of current, the DW can be expelled from the notch under a sufficiently large static magnetic field. Since the non-adiabatic contribution enters in the magnetization dynamics equation as an effective magnetic field\cite{Tatara2004,Zhang2004,Thiaville2005,Tatara2008,Boulle2011,Martinez2007b,Leliaert2015}, the STT parameters ($P$ and $\xi$) can be inferred by measuring the reduction or the enhancement of the required field as a function of the injected current. The experimental results of critical depinning current as a function of the applied field (or vice versa) are usually described in terms of standard micromagnetic and/or simple $1D$ models, which describe the DW dynamics governed by the Landau-Lifshitz-Gilbert (LLG) equation including the adiabatic and nonadiabatic STTs. However, these conventional approaches assume that the current density is flowing uniformly along the strip\cite{Martinez2007a,Martinez2009,Lepadatu2009}. This constitutes an oversimplification when the strip cross section is non-uniform, and especially when the current is forced to pass through constrictions.

\gg{Apart from the STTs, it is well known that an electric current also generates Joule heating. Although several experimental works have indicated that its effect can be significant~\cite{Hayashi2006a,Yamaguchi2005,Curiale2012,Emori2015,Togawa2008,Yamaguchi2012,Yamaguchi2006}, its influence on DW depinning has not yet been assessed. Reductions in the DW propagation field~\cite{Yamaguchi2006} or in the switching field~\cite{Togawa2008} have been experimentally observed in Py samples and ascribed to Joule heating. Furthermore, numerical studies\cite{Fangohr2011,Torrejon2012,Ramos2015} of the heat transport in these systems have shown that non-uniform temperature profiles can appear during the current pulse and that, depending on the current \gg{and the substrate}, the local temperature can be close to or even above the Curie temperature ($T_{C}$)}. Although thermal fluctuations at uniform room temperature have been studied numerically with the stochastic version of the Landau-Lifshitz-Gilbert equation\cite{Martinez2007a,Martinez2007b,Leliaert2015}, this Langevin approach is restricted to uniform temperatures well below the Curie threshold, and therefore it fails to describe the current-assisted DW depinning in systems where the temperature varies significantly. Here we develop a micromagnetic framework that couples both the electric and the heat transport to the magnetization dynamics self-consistently. Within this formalism, the magnetization dynamics is described by the Landau-Lifshitz-Bloch (LLB) equation\cite{Garanin1997,Chubykalo-Fesenko2006,Kazantseva2008,Atxitia2007,Hinzke2011,Schieback2009,Evans2012}, which allows us to properly describe the current-assisted DW depinning for temperatures close to or even above $T_{C}$.

\begin{figure}[h]
\includegraphics[width=0.45\textwidth]{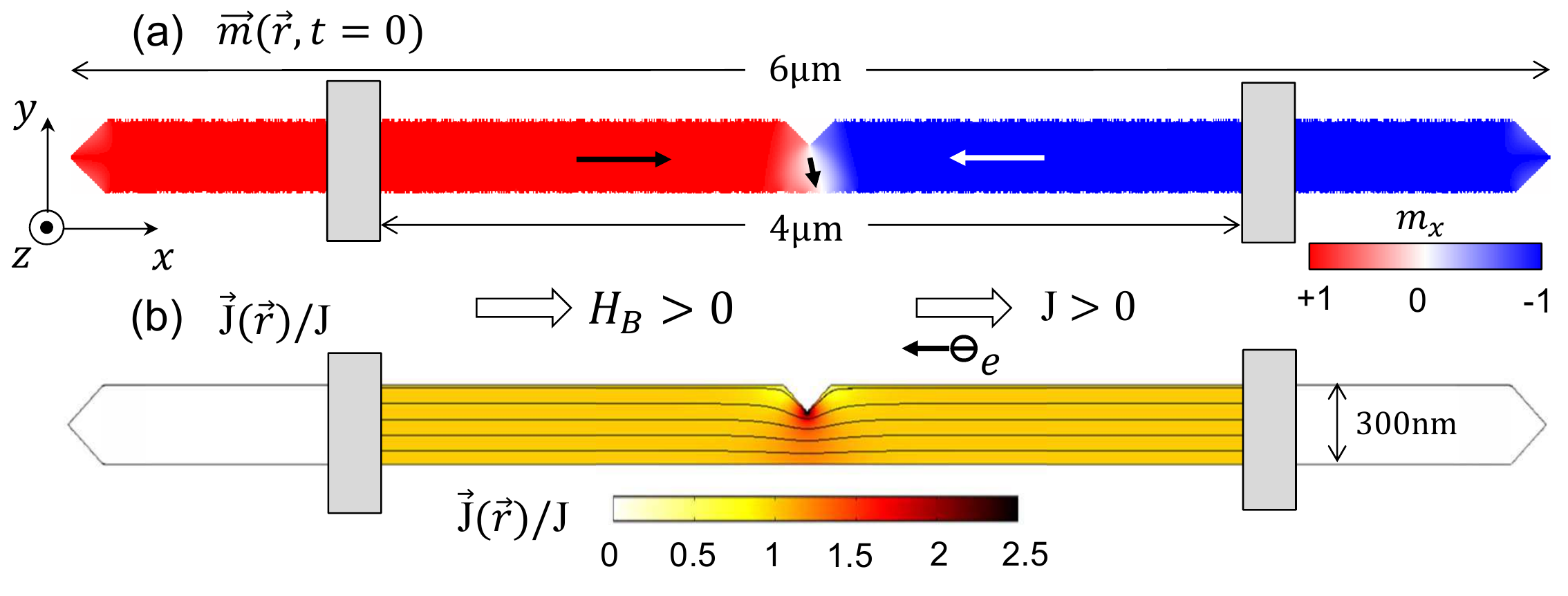}
\caption{(color on-line) (a) Geometry of the Py strip and initial magnetization state, $\vec{m}(\vec{r},t=0)$. (b) Spatial distribution of the current density $\vec{J}(\vec{r},t)$ during the current pulse $0\leq t \leq 4\mathrm{ ns}$. Lines represent schematically the local direction of the current density while the color represents its local module normalized to the value at points distant from the notch ($J$), where the current density is uniform across the strip width and points along the $x$-axis. }
\label{fig:Fig1}
\end{figure}

\indent As an archetypal experiment, the developed formalism is used here to numerically evaluate the phase diagram for current-assisted DW depinning studied experimentally by Hayashi et al.\cite{Hayashi2006a}. This experiment was performed on a $6\mathrm{\mu m}$-long Py nanostrip on top of a ${\rm SiO_2/Si}$ substrate. The Py strip has a $300\mathrm{nm}\times 10\mathrm{nm}$ cross section and contains a triangular notch with a depth of $\approx 100\mathrm{nm}$. Four different initial magnetic DW configurations pinned at the notch were observed, depending on the nucleation process: vortex or transverse DWs, pinned at the center or at the left side of the notch (see Fig. 2 in\cite{Hayashi2006a} for a detailed description of the nucleation process). After nucleation and pinning of DWs, $4\mathrm{ns}$ current pulses with both polarities ($J>0$ and $J<0$) of different amplitudes were applied through two gold contacts placed at a distance of $4\mathrm{\mu m}$ from each other under bias static fields $\vec{H}_{B}= H_{B}\vec{u}_x$ along the longitudinal $x$-axis. A sketch of the system is shown in Fig.~\ref{fig:Fig1}(a). After each current pulse, the presence or absence of the DW between the contacts was detected by resistance measurements, and the probability of current-assisted DW depinning was obtained as a function of the external field. The critical depinning current was defined as the one at which depinning probability exceeds $50\%$. As expected, in the absence of current pulses ($J=0$), the DWs initially trapped at the center of the notch were symmetrically depinned by positive ($H_{B}>0$) and negative ($H_{B}<0$) fields with the same magnitude. By contrast, the DWs pinned at the left side of the notch are more easily depinned by negative fields than by positive fields, simply because of the asymmetric pinning potential\cite{Hayashi2006a}. However, in the presence of current pulses ($J\neq 0$), two noticeable results can be observed in the measured depinning diagrams shown in Fig. 3 of Ref.\cite{Hayashi2006a}. First, DW depinning is not achieved in the absence of an applied field, and therefore, the evaluated currents alone do not exert a significant force on the DWs. Second, regardless of the initial magnetic configuration, all phase diagrams show a marginal dependence on the current polarity, i.e. positive ($J>0$) and negative ($J<0$) currents produce roughly the same phase diagram. These observations are not fully compatible with pure STT contributions\cite{Zhang2004,Thiaville2005}, which push the DW in opposite directions, depending on current polarity. By contrast, they suggest that the experimental results could be compatible with Joule heating (JH), which does not depend on the current polarity. In fact, Hayashi et al.~\cite{Hayashi2006a,Hayashi2006b} estimated an averaged temperature through the sample of about $\approx 780\mathrm{ K}$ during the injection of the pulses with $J\approx 3\times 10^8 \mathrm{A/cm^{2}}$. These temperatures are close to the Curie temperature of Py ($T_C\approx 850\mathrm{ K}$\cite{Yamaguchi2005}), underscoring the need to consider the effects of Joule heating.

\indent In this paper, we focus on the analysis of the current-assisted DW depinning from a notch under static fields and short current pulses, with emphasis on elucidating the effect of Joule heating. The paper is organized as follows. The conventional micromagnetic formalism based on the LLG equation and neglecting Joule heating is presented in Sec. \ref{sec:DWdepinning_NoJouleHeating}, which also includes its predictions and stresses its limitations to explain the experimental observations. In Sec.~\ref{sec:heat_dynamics} we describe the heat dynamics by means of electro-thermal simulations of the entire sample. A phenomenological model describing the heat transport is also presented and validated to further describe the heat transport coupled to the magnetization dynamics in the Py strip. The model developed for the heat transport is then coupled to the magnetization dynamics given by the LLB eq. in Sec. \ref{sec:DWdepinning_JouleHeating}, and the predictions for the DW depinning results are presented and compared with the experimental measurements. \b{Sec.~\ref{sec:discussion} presents a brief discussion of the results while the main conclusions of our study and their implications are summarized in Sec.~\ref{sec:conclusions}.}

\section{Domain wall depinning as predicted by the standard LLG model in the absence of Joule heating}
\label{sec:DWdepinning_NoJouleHeating}

In what follows, we focus our attention on the case of a transverse DW initially pinned in the center of the notch as shown in Fig. \ref{fig:Fig1}(b). Although the geometry and the dimensions of the Py strip were selected here to mimic the experimental samples studied by Hayashi et al.\cite{Hayashi2006a}, the exact geometry and dimensions of the notch may slightly differ. In the present study, a $100\mathrm{ nm}$-depth and $200\mathrm{ nm}$-wide triangular notch is assumed. In addition to the notch, realistic conditions were considered by taking into account roughness~\cite{Martinez2007a} at the edges, with a characteristic size $\approx 5\mathrm{ nm}$. In order to show the limitations of neglecting Joule heating, we first report the results obtained by standard micromagnetic simulations, where the magnetization dynamics at uniform and finite temperature $T\ll T_C$ is described by the Landau-Lifshitz-Gilbert (LLG) equation augmented by adiabatic and non-adiabatic STTs~\cite{Zhang2004,Martinez2007b}

\begin{eqnarray}
\frac{\partial\vec{m}}{\partial t} &=& - \gamma_0' \vec{m}\times\left( \vec{H}_{eff}+\vec{H}_{th}\right) \nonumber\\
&& - \gamma_0' \alpha\vec{m}\times\left( \vec{m}\times(\vec{H}_{eff}+\vec{H}_{th} )\right) \nonumber\\
&& - u(1+\xi\alpha)\vec{m}\times \left(\vec{m}\times (\vec{J}\cdot\nabla)\vec{m}\right)    \nonumber\\
&& - u(\xi-\alpha)\vec{m}\times(\vec{J}\cdot\nabla)\vec{m}\, ,\label{eq:LLG}
\end{eqnarray}

\noindent where $\vec{m}(\vec{r},t)=\vec{M}(\vec{r},t)/M_s$ is the normalized magnetization and $\vec{H}_{eff}(\vec{r},t)$ is the effective field, which includes the standard micromagnetic contributions~\cite{Martinez2007b,LopezDiaz2012}. $\gamma_0'=\gamma_0/(1+\alpha^2)$ where $\alpha$ is the Gilbert damping and $\gamma_0$ the gyromagnetic ratio. $\vec{J}(\vec{r},t)$ is the current density and $u=\mu_{B}P^{0}/(M_{s}^{0} e(1+\xi^2))$, where $\mu_B$ is the Bohr magneton and $e$ the elementary charge. $P^0$ and $M_{s}^{0}$ are the current polarization and the saturation magnetization at zero temperature, and $\xi$ represents the dimensionless non-adiabatic parameter~\cite{Zhang2004,Thiaville2005,Tatara2008}. Thermal fluctuations at uniform and constant room temperature $T=300\mathrm{ K}$ are taken into account through a stochastic thermal field $\vec{H}_{th}$, which has white noise properties with the correlator~\cite{Martinez2007b,Brown1963,GarciaPalacios1998}

\begin{eqnarray}
\langle H_{th}^i(\vec{r},t)H_{th}^j(\vec{r}',t')\rangle=\frac{2\alpha k_BT}{\gamma_0\mu_0M_sV}\delta_{ij}\delta(\vec{r}-\vec{r}')\delta(t-t')  ,\nonumber\\
\end{eqnarray}

\noindent where $k_B$ is the Boltzmann constant, $\mu_0$ the vacuum permeability and $V$ the volume of the computational cell. Starting from the initial state depicted in Fig. \ref{fig:Fig1}(a), $4\mathrm{ns}$ current pulses together with \r{a bias field $\vec{H}_{B}=H_{B}\vec{u}_{x}$} are applied and the magnetization dynamics is studied by numerically solving Eq.~(\ref{eq:LLG}) using a $4$th-order Runge-Kutta scheme. 
$\vec{J}(\vec{r},t)$ was previously calculated by COMSOL\cite{comsol} (see Fig. \ref{fig:Fig1}(b)) and included in Eq.~(\ref{eq:LLG}). Typical Py parameters were considered: $M_s\equiv M_s^0=8.6\times 10^5\ {\rm A/m^2}$, $A^0=1.3\times 10^{-11}\ {\rm J/m}$ (exchange constant at zero temperature), $\alpha=0.02$, $P^{0}=0.4$ and $\xi=0.04$~\cite{Boulle2011}. The micromagnetic results described below were obtained by using a computational time step of $\Delta t= 0.1\mathrm{ ps}$, and it was checked in several tested cases that a reduced time step of $\Delta t= 0.05\mathrm{ ps}$ did not modify the obtained results. The Py strip is spatially discretized using cubic computational cells $\Delta x=\Delta y=\Delta z =5\mathrm{ nm}$. In order to obtain the depinning probability at $T=300\mathrm{K}$, $10$ stochastic realizations were evaluated. Under these conditions, we computed the critical DW depinning current $J_d$ as a function of the bias field. As in Ref.~\cite{Hayashi2006a}, $J_d$ is defined as the minimum current density needed to depin the DW with a probability higher than $50\%$. Henceforth, we shall use $J$ to indicate the module of the current density at points distant from the notch, where $\vec{J}$ is uniform across the strip width and points along the longitudinal direction ($\vec{u}_{J}=\vec{u}_{x}$). Note that, due to the notch, $|\vec{J}|$ is higher below the notch and has a $y$ component (Fig.~\ref{fig:Fig1}(b)). The DW was considered as depinned when it was expelled outside the gold contacts (Fig. \ref{fig:Fig1}), as would be obtained with magneto-resistance measurements\cite{Hayashi2006a}.

\begin{figure}[h]
\includegraphics[width=0.4\textwidth]{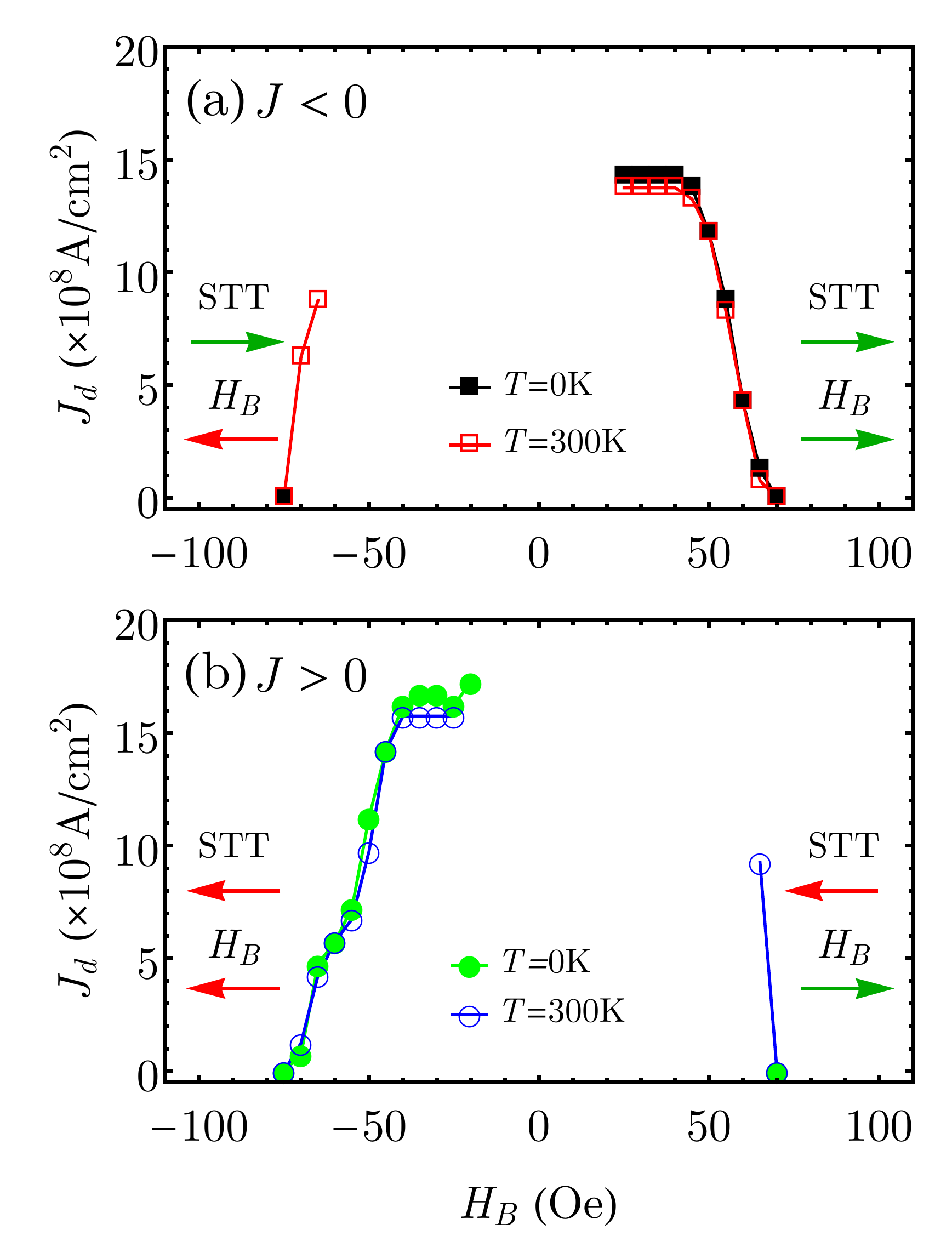}
\caption{(color on-line) Depinning current ($J_d$) as a function of the bias field ($H_B$) obtained with the standard LLG equation in the absence of Joule heating for (a) negative and (b) positive current pulses. Results are shown for $0$ and constant room temperature. ${\rm STT}$ (${H_B}$) pushes the DW in the direction of the arrows. 
Errors bars, corresponding to the adopted current step, are smaller than the plot markers.}
\label{fig:Fig2_PhaseD_LLG}
\end{figure}

\indent The results are shown in Fig.~\ref{fig:Fig2_PhaseD_LLG}(a) and \ref{fig:Fig2_PhaseD_LLG}(b) for negative and positive currents respectively. In the absence of current ($J=0$), the depinning fields are $H_{d}^{\mp}=-75/+70\ {\rm Oe}$; i.e., slightly higher than the experimental ones~\cite{Hayashi2006a} ($H_{d}^{\mp}=-54/+54\ {\rm Oe}$), meaning a slightly different notch. From the results collected in Fig.~\ref{fig:Fig2_PhaseD_LLG}, two noticeable differences with the experimental observations (see Fig. 3(a) in \cite{Hayashi2006a}) can be found: On the one hand, the amplitude of the depinning currents $J_{d}$ is around $5$ times higher than the ones measured experimentally. On the other hand, the experimental depinning diagrams depicted in Fig.~3(a) of Ref.~\cite{Hayashi2006a} are almost independent of the current polarity, i.e., positive and negative currents produce roughly the same reduction on the depining field. However, and in agreement with the STT contribution that pushes the DW in opposite directions, here we found an evident asymmetric depinning diagram with respect to the current polarity. \b{For instance, under negative current pulses ($J<0$ with the electron flowing along the positive $x$ direction, Fig.~\ref{fig:Fig2_PhaseD_LLG}(a)), we found more depinning events for $H_B>0$, where the STT and the bias field push the DW in the same direction. If we reverse the current ($J>0$, Fig.~\ref{fig:Fig2_PhaseD_LLG}(b)) we have more depinning events for $H_B<0$. 
This asymmetric behaviour is due to the balance between the two driving forces: while for $J<0$ the STT pushes the DW to the right, the driving field pushes the DW towards the right or left depending on its direction.} 
We checked that it was not possible to reproduce the experimental phase diagram for any of the combinations of the STT parameters ($0\le P^0 \le 1$ and $0\le\xi\le 5\alpha$): an increase in $\xi$ decreases the depinning current but increases the asymmetry with the current polarity. Also, an increase of $P^0$ up to unrealistic values ($P^0=1$) can decrease the depinning current but does not lead to the experimentally observed symmetry. These results indicate that conventional micromagnetic simulations are inadequate to reproduce the experimental results in these systems. As mentioned, Hayashi et al.~\cite{Hayashi2006a} already estimated an average temperature between the contacts of about $T_{av}\approx 780\mathrm{ K}$ during the injection of the pulses with $J\approx 3\times 10^{8} \mathrm{A/cm^{2}}$. 
These temperatures, being close to the Curie temperature of Py ($T_C\approx 850\mathrm{ K}$), suggest that a more realistic description, that takes into account Joule heating effects, needs to be adopted in order to correctly describe the current-assisted DW depinning in these systems. 

\section{Heat dynamics}
\label{sec:heat_dynamics}

In order to evaluate the relevance of Joule heating effects on the current-assisted DW depinning, a preliminary analysis of heat transport is required. The temperature evolution of the sample is governed by the heat equation

\begin{equation}
\frac{\partial T(\vec{r},t)}{\partial t}=\frac{k}{\rho C}\nabla^2 T(\vec{r},t)+\frac{Q(\vec{r},t)}{\rho C}\, ,
\label{eq:heat_equation_first}
\end{equation}

\noindent where $k$ is the thermal conductivity, $\rho$ the density, and $C$ the specific heat capacity of the material. $T=T(\vec{r},t)$ represents the temperature and $Q$ the heat source. For Joule heating $Q(\vec{r},t)=\vec{J}(\vec{r},t)^2/\sigma$, with $\vec{J}(\vec{r},t)$ the current density, and $\sigma$ the electrical conductivity. \gg{Analytical solutions of Eq.~(\ref{eq:heat_equation_first}) have been obtained under the assumption of uniform temperature and semi-infinite substrate~\cite{You2006,Kim2008}.} \gg{Here, due to the non-uniform source and the finite substrate thickness, we solved Eq.~(\ref{eq:heat_equation_first}) for the full system ($\mathrm{Py/SiO_{2}/Si}$) by COMSOL simulations.}  Typical Py parameters~\cite{Fangohr2011} were considered: $\rho_{Py}=8.7\times10^3\ \rm{Kg/m}^3$, $C_{Py}=0.43\ \rm{ J/(gK)}$, $k_{Py}=46.4\ \rm{ W/(Km)}$, $\sigma_{Py}=4\times 10^6\ (\rm{ \Omega m})^{-1}$. For the $\rm{Si}$ substrate, nominal parameters were also assumed: $\rho_{\rm{Si}}=2.329 \times10^3\ \rm{ Kg/m}^3$, $C_{\rm{Si}}=0.7\ \rm{J/(gK)}$ and $k_{\rm{Si}}=130\ \rm{W/(Km)}$. The presence of a thin native $\rm{SiO_2}$ interlayer~\cite{Hayashi2006b} was also taken into account. Consistent with a previous analysis\cite{Ramos2015}, this interlayer imposes a thermal resistance\cite{Cahill2003} between the Py strip and Si substrate (Ref.~\cite{comsol}). Here a thermal resistance of $2.2 \times 10^{-8} \rm{ m^2 K/W}$ was adopted to reproduce the experimental results\cite{Hayashi2006a}. Owing to the high thermal conductivity of Au, the electrical contacts are considered to be sinks for the heat, and therefore their temperature is fixed to room temperature. The electrical conductivity is assumed to be different from zero only in the Py and in the Au contacts.

\begin{figure}[h]
\includegraphics[width=0.4\textwidth]{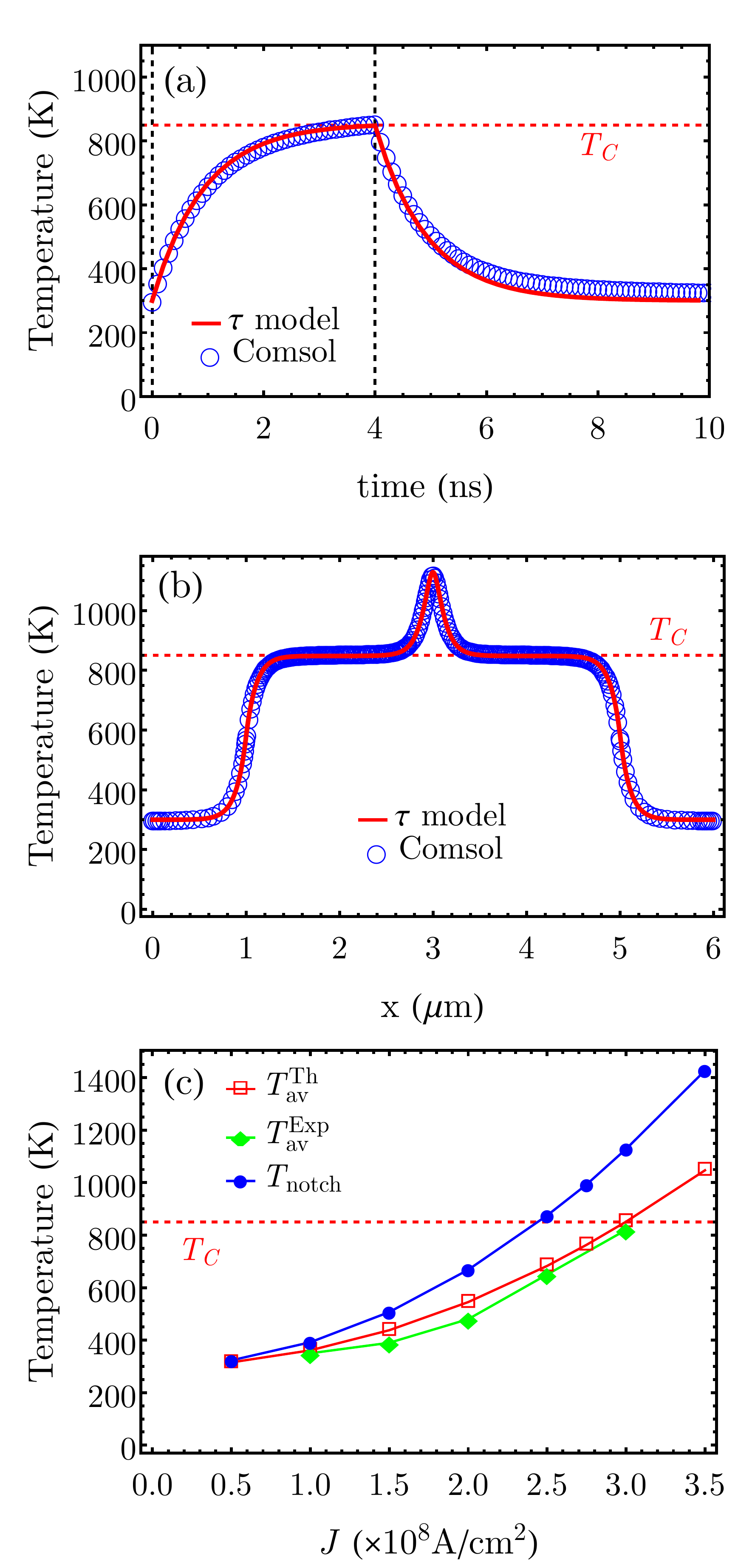}
\caption{(color on-line) (a) Temporal evolution of the averaged temperature ($T_{av}^{Th}(t)$) in the Py strip for $J=3\times10^{8}\rm{A/cm^2}$. The average was computed between the two contacts used to inject the current pulse. (b) Temperature profile ($T(x,y=150\mathrm{nm})$ $vs$ $x$) in the middle of the strip ($y=150\mathrm{nm}$) at $t=4\mathrm{ ns}$ for the same current $J=3\times 10^{8}\rm{ A/cm^2}$. (c) Average temperature ($T_{av}^{Th}$) at the end the pulse ($t=4\mathrm{ns}$) as a function of the applied current density $J$. The experimental data ($T_{av}^{Exp}$) were extracted from Ref.~\cite{Hayashi2006a}. The theoretically computed local temperature at the notch location ($T_{notch}$) is also shown.}
\label{fig:Fig3_FigureT}
\end{figure}

As an example, the temperature of the Py strip is shown for a $4\mathrm{ns}$ current pulse with a magnitude of $J=3\times10^{8}\rm{ A/cm^2}$. COMSOL predictions are shown by blue dots in Fig. \ref{fig:Fig3_FigureT}. The temporal evolution of the Py temperature, averaged over the region between the Au contacts ($T_{av}^{Th}$), is shown in Fig. \ref{fig:Fig3_FigureT}(a). Upon application of the current pulse, $T_{av}^{Th}$ increases monotonously (see Fig. \ref{fig:Fig3_FigureT}(a) for $0<t<4\mathrm{ ns}$), and it relaxes again to room temperature once the current pulse is switched off (see Fig. \ref{fig:Fig3_FigureT}(a) for $t>4\mathrm{ ns}$). The local temperature profile ($T(\vec{r},t)$) along the middle line of the Py strip ($y=150\mathrm{ nm}$), at the end of the pulse ($t=4\mathrm{ ns}$), is shown in Fig. \ref{fig:Fig3_FigureT}(b). This picture clearly indicates that the temperature is not uniform along the Py strip, with remarkable variations around the notch, where the current density is higher (see Fig. \ref{fig:Fig1}(b)).

\indent Since the inclusion of the substrate into the micromagnetic code would be prohibitive from a computational point of view~\cite{note5}, we adopted a phenomenological model of the heat transport by adding an additional convective term into Eq.~(\ref{eq:heat_equation_first}), which now reads\cite{Perez2015}

\begin{equation}
\frac{\partial T(\vec{r},t)}{\partial t}=\frac{k}{\rho C}\nabla^2 T(\vec{r},t)+\frac{Q(\vec{r},t)}{\rho C}-\frac{T(\vec{r},t)-T_{0}}{\tau}\, 
\label{eq:heat_equation}
\end{equation}

\noindent where the last term in Eq.~(\ref{eq:heat_equation}) takes into account the cooling of the wire through the substrate and the surrounding ambient. The parameter $\tau$ represents the characteristic time rate at which the heat flows from the strip to the substrate and it was fixed to $\tau=0.9\mathrm{ns}$ in order to have, for all the currents and at the end of the $4$ ns pulse, the same averaged temperature as estimated experimentally in Ref.~\cite{Hayashi2006a} ($T_{av}^{Exp}$), and also obtained by the COMSOL simulations described above. Unlike Eq.~(\ref{eq:heat_equation_first}), the phenomenological heat Eq.~(\ref{eq:heat_equation}) is only solved for the Py strip, and therefore, it can be implemented to evaluate the heat transport and the magnetization dynamic simultaneously. In order to validate this phenomenological model, its predictions are depicted in Fig.~\ref{fig:Fig3_FigureT}(a) and (b) by solid red lines. Both the temporal evolution (Fig.~\ref{fig:Fig3_FigureT}(a)) and the local temperature profile (Fig.~\ref{fig:Fig3_FigureT}(b)) are in remarkable agreement with COMSOL results, which allows us to easily couple the heat transport to the magnetization dynamics in our micromagnetic code.

Fig.~\ref{fig:Fig3_FigureT}(c) shows the Py temperature as a function of $J$. The theoretical average temperature ($T_{av}^{Th}$) is in agreement with the experimental values ($T_{av}^{Exp}$) of Ref.~\citep{Hayashi2006a}. Note that for high current pulses ($J\ge 2.5\times 10^{8}\rm{ A/cm^2}$), even if the average temperature $T_{av}$ remains below the Curie point, the local temperature at the notch ($T_{notch}$) reaches values above $T_C$ (Fig.~\ref{fig:Fig3_FigureT}(c)), leading to a local destruction of the ferromagnetic order. \gg{In agreement with previous analysis~\cite{Fangohr2011} we found that the temperature dynamics strongly depends on the substrate. In particular, in our case it depends on the thickness of the small $\rm{SiO_2}$ inter-layer. This can explain the comparable increases in temperature observed in similar Py samples for much lower current densities~\cite{Togawa2008, Yamaguchi2012,Yamaguchi2006}. In some of those systems for instance, the $\rm{SiO_2}$ insulating layer was much thicker ($\approx 100$ nm)~\cite{Yamaguchi2012,Yamaguchi2006} or the substrate was different~\cite{Togawa2008,Yamaguchi2006} leading to a less efficient heat absorption and a higher increase in temperature.}

\section{Coupled magneto-heat dynamics: the influence of Joule heating}
\label{sec:DWdepinning_JouleHeating}

\begin{figure}[h]
\includegraphics[width=0.45\textwidth]{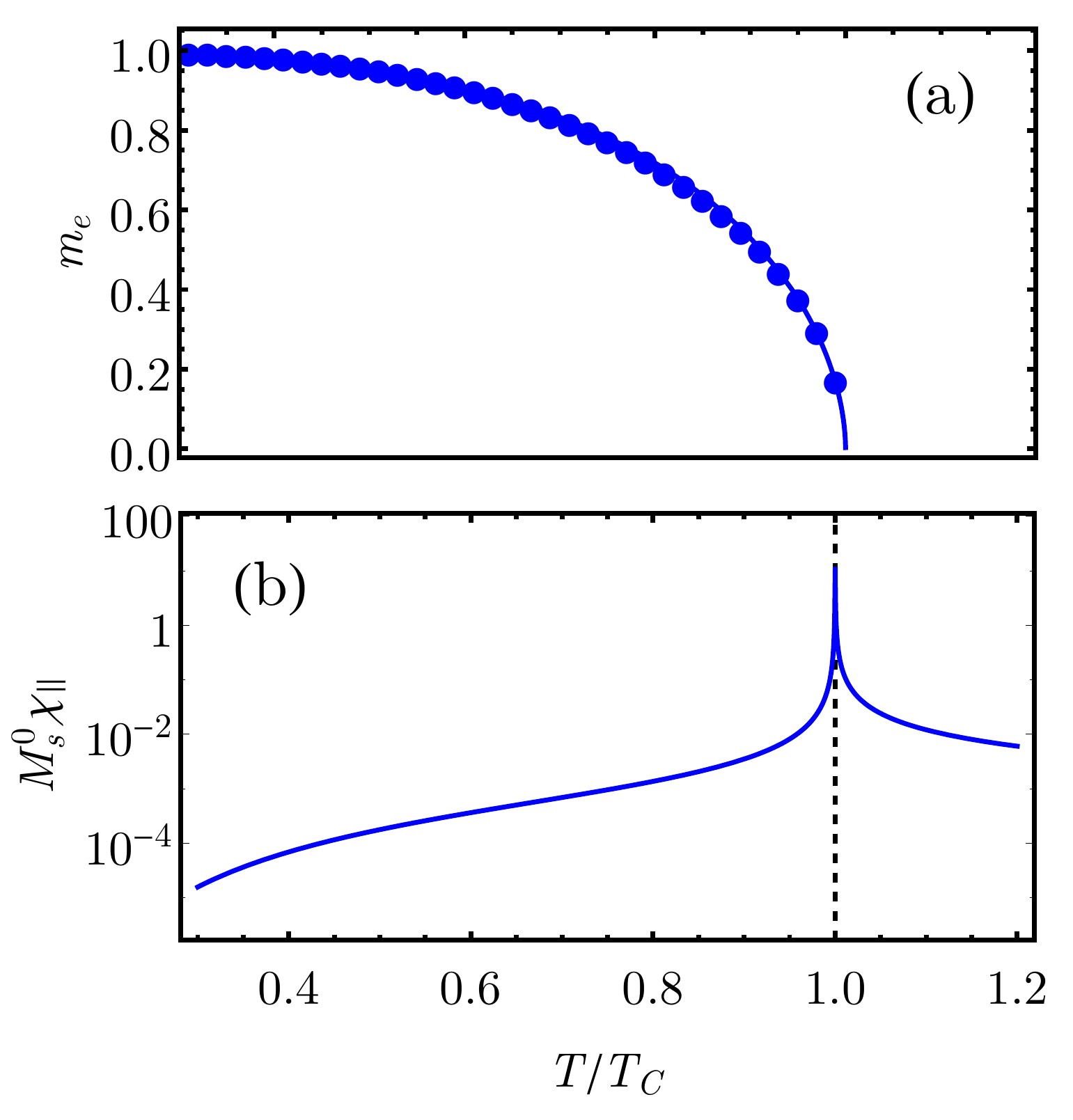}
\caption{\gg{(color on-line) (a) Normalized equilibrium magnetization $m_e(T)$ and (b) longitudinal susceptibility $\chi_{\|} (T)$ multiplied by $M_s^0$ (in order to be dimensionless) as function of temperature. Dots in (a) represent a numerical solution of Eq.~(\ref{eq:me}). The line is a fit of the solution.}}
\label{fig:Fig4bis_mechi}
\end{figure}

We showed in Sec.~\ref{sec:DWdepinning_NoJouleHeating} that the analysis of the current-assisted DW depinning in the framework of the conventional micromagnetic model, which includes the STTs but neglects Joule heating phenomena, is not sufficient to explain several features observed in the experiment\cite{Hayashi2006a}. In particular, the model overestimates the magnitude of the critical depinning currents as compared to those observed in the experiment. Moreover, the lack of symmetry with respect to the current polarity in the depinning phase diagrams predicted by this oversimplified model is not consistent with the experimentally observed symmetry. 
Additionally, both the experimental measurements of the average temperature  and the theoretical analysis performed in Sec. \ref{sec:heat_dynamics}, show that Joule heating needs to be taken into account in order to provide a more realistic description of these systems. Therefore, the magnetization dynamics must be studied as being coupled to the heat transport excited by the current pulses. Since the sample reaches temperatures near or even above $T_C$ we can not use the LLG equation, valid for $T\ll T_C$. A natural choice is the Landau-Lifshitz-Bloch (LLB) equation~\cite{Garanin1997}, which can describe the magnetization dynamics for a wide range of temperatures, even above $T_C$~\cite{Chubykalo-Fesenko2006,Kazantseva2008}. It has been used to describe ultrafast demagnetization processes~\cite{Atxitia2007} as well as DW motion by thermal gradients\cite{Hinzke2011}. Here we extend its application to the analysis of the JH effect in current-assisted DW depinning. The STT contributions were included in the LLB eq.~\cite{Evans2012} by Schieback et al.~\cite{Schieback2009}. The LLB equation describing the magnetization dynamics under STT reads as

\begin{eqnarray}
\frac{\partial\vec{m}}{\partial t}&=&-\gamma_0'(\vec{m}\times \vec{H}_{eff})+\gamma_0'\frac{\alpha_{\|}}{m^2}\left(\vec{m}\cdot\vec{H}_{eff}\right)\vec{m}  \nonumber\\
&&-\gamma_0'\frac{\alpha_{\perp}}{m^2}\left(\vec{m}\times\left(\vec{m}\times (\vec{H}_{eff}+\vec{H}_{th}^{\perp})\right)\right)+\vec{H}_{th}^{\|} \nonumber\\
&& + u\left(1+\frac{\xi\alpha_{\perp}}{m}\right)(\vec{J}\cdot\nabla)\vec{m}         \nonumber\\
&& - \frac{u}{m}\left(\xi-\frac{\alpha_{\perp}}{m}\right)\vec{m}\times(\vec{J}\cdot\nabla)\vec{m} \nonumber\\
&& - \frac{u\xi\alpha_{\perp}}{m^{3}}\left( \vec{m} \cdot (\vec{J}\cdot\nabla)\vec{m}\right)\vec{m}\, ,
\label{eq:LLB}
\end{eqnarray}

\noindent where $\gamma_0'=\gamma_0/(1+\lambda^2)$. \b{Apart from the conventional precessional and damping terms (first and third term on the RHS of Eq.~\ref{eq:LLB}) the LLB equation includes an additional longitudinal relaxation term (second term on the RHS  of Eq.~\ref{eq:LLB}) which describes the relaxation of the module of $\vec{m}$ towards its equilibrium value $m_e(T)$. \gg{$m_e(T)$ represents the normalized equilibrium magnetization at each temperature, which we calculated by using the Brillouin function for $1/2$ spins, namely~\cite{Bertotti,note4}
\begin{equation}
m_e=\tanh\left[\frac{\mu_0\mu_{\rm Py}}{k_B T}H_B+\frac{T_{C}}{T}m_e\right]\, ,
\label{eq:me}
\end{equation} 
where $\mu_{\rm Py}$ represents the Py atomic magnetic moment and $H_B$ the external field. We assumed $\mu_{\rm Py}=\mu_B$ according to previous calculations~\cite{Mijnarends}}. A characteristic feature of the LLB is precisely that, contrary to the LLG,  $\vec{m}$ is not unitary but its module varies depending on the temperature. Note that, this longitudinal relaxation is neglected within the LLG formalism since, at $T=300 K$, it is much faster than the transverse relaxation. However it becomes particularly important at $T$ close to $T_C$ where longitudinal and transverse relaxation times are comparable. 
$\alpha_{\perp}$ and $\alpha_{\|}$ are the transverse and the longitudinal damping parameters respectively. They depend on the temperature as~\cite{Garanin1997,Chubykalo-Fesenko2006,Kazantseva2008} $\alpha_{\perp}=\lambda(1-T/(3T_C))$ and $\alpha_{\|}=2\lambda T/(3T_C)$ for $T<T_C$, while $\alpha_{\perp}=\alpha_{\|}=2\lambda T/(3T_C)$ for $T>T_C$. $\lambda$ represents a microscopic damping parameter, and in the limiting case of zero temperature $\alpha_{\perp}$ reduces to the conventional Gilbert damping with $\alpha\equiv\lambda$ and $\alpha_{\|}=0$.} The effective field, $\vec{H}_{eff}$, is given by~\cite{Garanin1997,Chubykalo-Fesenko2006,Kazantseva2008}

\begin{eqnarray}
\vec{H}_{eff}&=&\vec{H}_{exch}+\vec{H}_{B}+\vec{H}_{dmg}\nonumber\\
&&+\left\{\begin{array}{ll}
\frac{1}{2\chi_{\|}}\left(1-\frac{m^2}{m_e^2}\right)\vec{m}, & T<T_C\\
-\frac{1}{\chi_{\|}}\left(1+\frac{3}{5}\frac{T_Cm^2}{(T-T_C)}\right)\vec{m},  & T>T_C\\
\end{array}\right.\, ,
\end{eqnarray}

\noindent where $\vec{H}_{B}$, $\vec{H}_{dmg}$ are the external bias field and the demagnetizing field respectively. $\chi_{\|}$ is the so-called longitudinal susceptibility, defined as $\chi_{\|}=\partial m_e/\partial H_B|_{H_B\rightarrow 0}$ ~\cite{Chubykalo-Fesenko2006,Kazantseva2008}, \gg{straightforward calculations give for $\chi_{\|}$:
\begin{equation}
\chi_{\|}=\left.\frac{\frac{\mu_0\mu_B}{k_BT}B'(x)}{1-\frac{T_C}{T}B'(x)}\right]_{x=\frac{T_C}{T}m_e}\, ,
\end{equation}
being $x=\frac{\mu_0\mu_B}{k_BT}H_B+\frac{T_C}{T}m_e$ and $B'(x)=\frac{\partial \tanh x}{\partial x}$. $m_e(T)$ and $\chi_{\|} (T)$ are shown in Fig.~\ref{fig:Fig4bis_mechi}}.

\noindent The exchange field $\vec{H}_{exch}$ is given by~\cite{Kazantseva2008,Schieback2009,Hinzke2011}

\begin{equation}
\vec{H}_{exch}=\frac{2A(T)}{m_e^2M_s^0}\nabla^2\vec{m}\, ,
\end{equation}

\noindent where $A(T)$ is the temperature-dependent exchange constant. Here we follow the same assumption adopted in Ref.~\cite{Atxitia2007,Ramsay2015}, where $A(T)$ scales with $T$ as $A(T)=A^0m_e^2(T)$. $M_s^0$ and $A^0$ represent the saturation magnetization and the exchange constant at $T=0$. 
The stochastic fields $\vec{H}_{th}^{\perp,\|}$ take into account longitudinal and transverse fluctuations of the magnetization. They are considered to have white noise properties with the following correlators~\cite{Evans2012}

\begin{eqnarray}
\langle H_{th}^{\perp i}(\vec{r},t)H_{th}^{\perp j}(\vec{r}',t)\rangle=&& \nonumber\\
=\frac{2k_BT(\alpha_{\perp}-\alpha_{\|})}{\gamma_0\mu_0M_s^0V\alpha_{\perp}^2}&\delta_{ij}&\delta(\vec{r}-\vec{r}')\delta(t-t')\, ,\nonumber\\
\langle H_{th}^{\| i}(\vec{r},t)H_{th}^{\| j}(\vec{r}',t)\rangle=&& \nonumber\\
=\frac{2\gamma_0k_BT\alpha_{\|}}{\mu_0M_s^0V}&\delta_{ij}&\delta(\vec{r}-\vec{r'})\delta(t-t')\, .
\label{eq:thfield}
\end{eqnarray}

The last three terms in Eq.~(\ref{eq:LLB}) describe the STT as introduced by Schieback et al.~\cite{Schieback2009,note1}. The temperature dependence of the current polarization is assumed to be~\cite{Schieback2009} $P=P^0m_e(T)$.  Eqs.~(\ref{eq:heat_equation}) and (\ref{eq:LLB}) are simultaneously solved by Heun's method for a $4\mathrm{ ns}$ current pulse over a total simulation time of $10\mathrm{ns}$. Note that within this magneto-thermal framework, the magnetization dynamics, eq. (\ref{eq:LLB}), is coupled to the temperature dynamics, eq. (\ref{eq:heat_equation}), both through the thermal fields and the temperature dependence of the magnetic parameters.

\begin{figure}[h]
\includegraphics[width=0.40\textwidth]{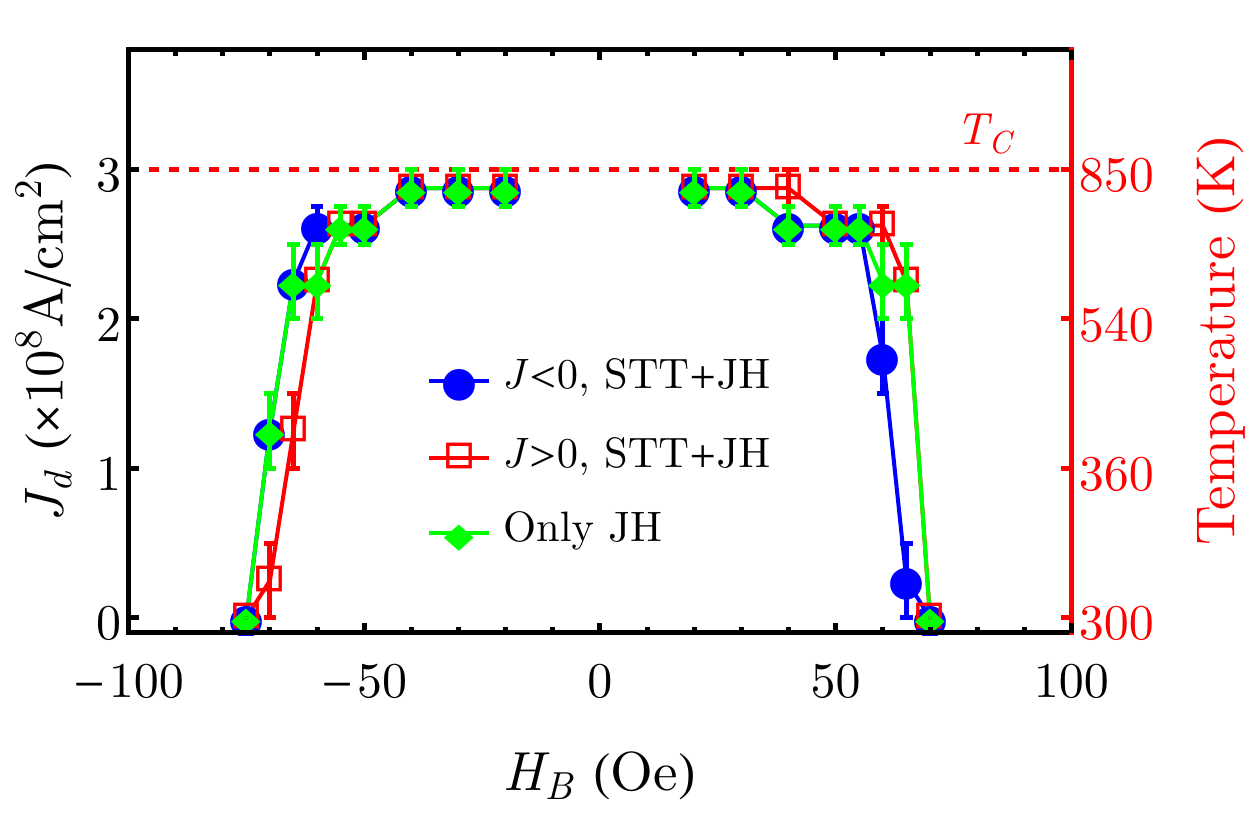}
\caption{(color on-line) Depinning current ($J_d$) vs bias field ($H_B$) obtained with the LLB eq. (\ref{eq:LLB}) including Joule heating (\ref{eq:heat_equation}). Results including the STTs are shown for positive (open red squares), \v{and} negative (full blue circle) currents. Green diamonds correspond to the results obtained in the absence of STTs ($P^0=0$). The scale (non linear) at the right-hand side shows the average temperature $T_{av}$ between the contacts for each current at $4\mathrm{ ns}$. }
\label{fig:Fig4_PhaseD_LLB}
\end{figure}

\indent The new depinning diagrams are shown in Fig.~\ref{fig:Fig4_PhaseD_LLB}. The depinning currents are now in quantitative agreement with the experimental measurements (see Fig. 3(a) in Ref.~\cite{Hayashi2006a}). Moreover, the symmetry with respect to the current polarity of the phase diagram is also reproduced. Indeed, the depinning currents are almost independent of the current polarity ($J>0$ or $J<0$). These observations can be explained by the pivotal role of Joule heating, which has no dependence on the current direction. The main effects of the temperature rise are the increase in thermal fluctuations, the decrease in the depinning field, and the local destruction of the ferromagnetic order for high currents. While the first two effects, together with the applied field and STT, are responsible for the points at $J<2.5\times10^8\ {\rm A/cm^2}$, the last one drives the depinning for $J\geq 2.75\times10^8\ {\rm A/cm^2}$.

\begin{figure*}[ht]
\includegraphics[width=1.0\textwidth]{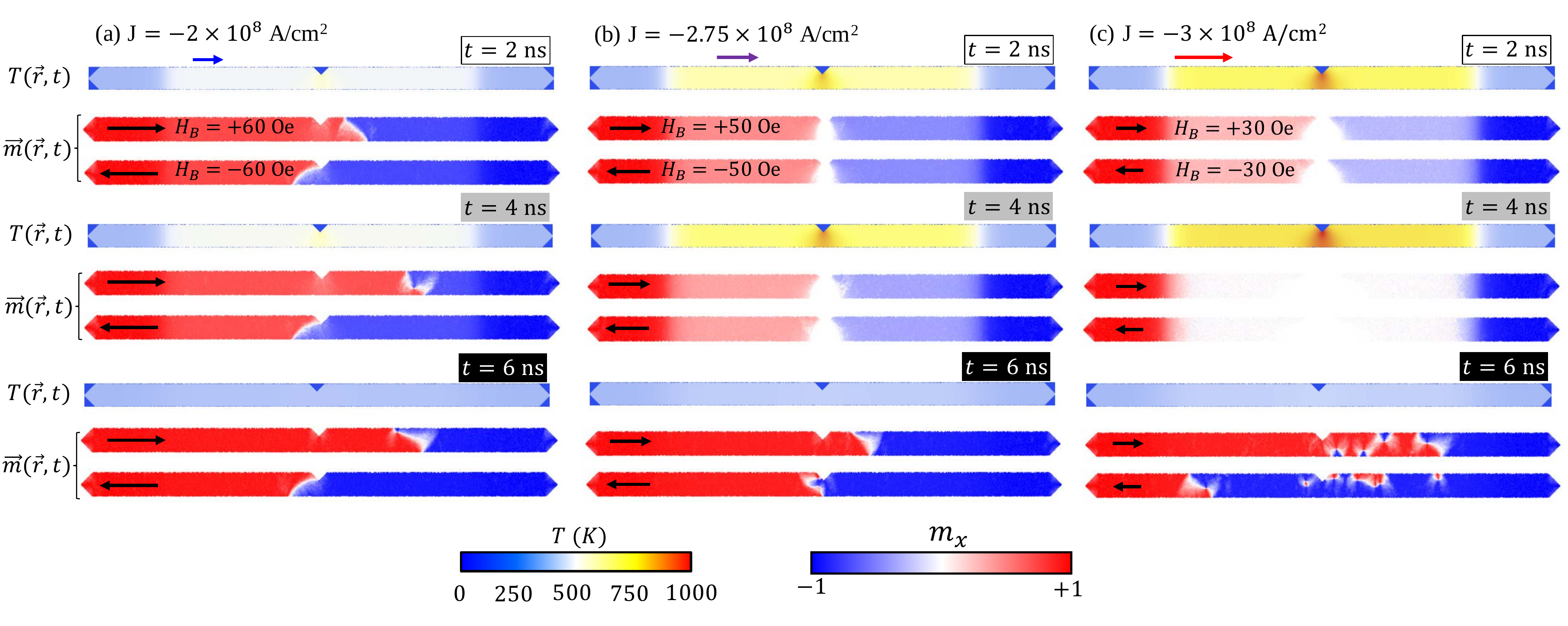}
\caption{(a) Magnetization and temperature dynamics for $J=-2\times10^8\ \rm{A/cm^2}$ (Multimedia view). (b)  Magnetization and temperature dynamics for $J=-2.75\times10^8\ \rm{A/cm^2}$ (Multimedia view). (c)  Magnetization and temperature dynamics for $J=-3\times10^8\ \rm{A/cm^2}$ (Multimedia view). \gg{Static images represent the temperature and magnetization patterns for the corresponding current and fields at $t=2$, $4$ and $6$ ns.} }
\label{fig:MoviesSnap}
\end{figure*}

The scale at the right hand side of Fig.~\ref{fig:Fig4_PhaseD_LLB} shows the average temperature $T_{av}^{Th}$ between the contacts for each current pulse at $4\mathrm{ns}$. 
The local temperature can deviate from this average, and, depending on $J$, it can reach higher values around the notch, where the current is higher (Fig. \ref{fig:Fig3_FigureT}(b)). Accordingly, the mechanism by which the DW is depinned depends on the magnitude of the current injected. \b{In order to visualize the magnetization dynamics together with the temperature evolution along the strip, several representative movies are available in Fig.~\ref{fig:MoviesSnap} (Multimedia view). The movies correspond to the cases $J=-2$ (Fig.~\ref{fig:MoviesSnap}(a)), $-2.75$ (Fig.~\ref{fig:MoviesSnap}(b)) and $-3\times10^8\ {\rm A/cm^2}$ (Fig.~\ref{fig:MoviesSnap}(c)), with different bias fields.} \gg{The static images in Fig.~\ref{fig:MoviesSnap} represent the temperature and magnetic patterns for the corresponding current and fields at $t=2$, $4$ and $6$ ns. It is possible to see the different magnetic dynamics depending on the current (temperature) and field. }

During the injection of a current pulse of $|J|=3\times10^8\ {\rm A/cm^2}$, the temperature between the contacts (not only under the notch) reaches $T_C$, and therefore the ferromagnetic order is temporarily destroyed between them (see Fig.~\ref{fig:MoviesSnap} (c)). Once the current pulse is switched off ($t>4\mathrm{ ns}$), the ferromagnetic order is recovered as $T(\vec{r},t)$ relaxes back to uniform room temperature, as described in Fig. \ref{fig:Fig3_FigureT}(a). Finally, a new DW is re-nucleated outside the notch. This re-nucleation process is influenced by the shape anisotropy and the static applied field. The ferromagnetic order is not destroyed outside the region between the contacts because no current has flowed there. These outside regions preserve a quasi-uniform magnetization pointing in opposite directions along the $x$-axis and, as a consequence, a DW is forced to be re-nucleated between the contacts for $t>4\mathrm{ns}$. Note that once the DW is out of the notch, it also undergoes a local pinning due to the edge roughness, which generates a minimum DW propagation field along the strip of $H_p\approx15\ {\rm Oe}$. Therefore, depending on the applied field, the re-nucleated DW can finally be expelled from the contact area, resulting in a depinning event. The absence of points for $|H_B|<20\ {\rm Oe}$ is precisely due to this propagation field. In agreement with the experimental results, no DW depinning is achieved in the absence of external field. \gg{The emergence of multiple-domains states, as consequence of re-nucleation due to Joule heating above $T_C$, was also observed experimentally in other studies on Py strips~\cite{Yamaguchi2012,Yamaguchi2006}. Here due to the reduced dimensions and the sides effect, we mainly observe one re-nucleated DW.}

\begin{figure}[h]
\includegraphics[width=0.45\textwidth]{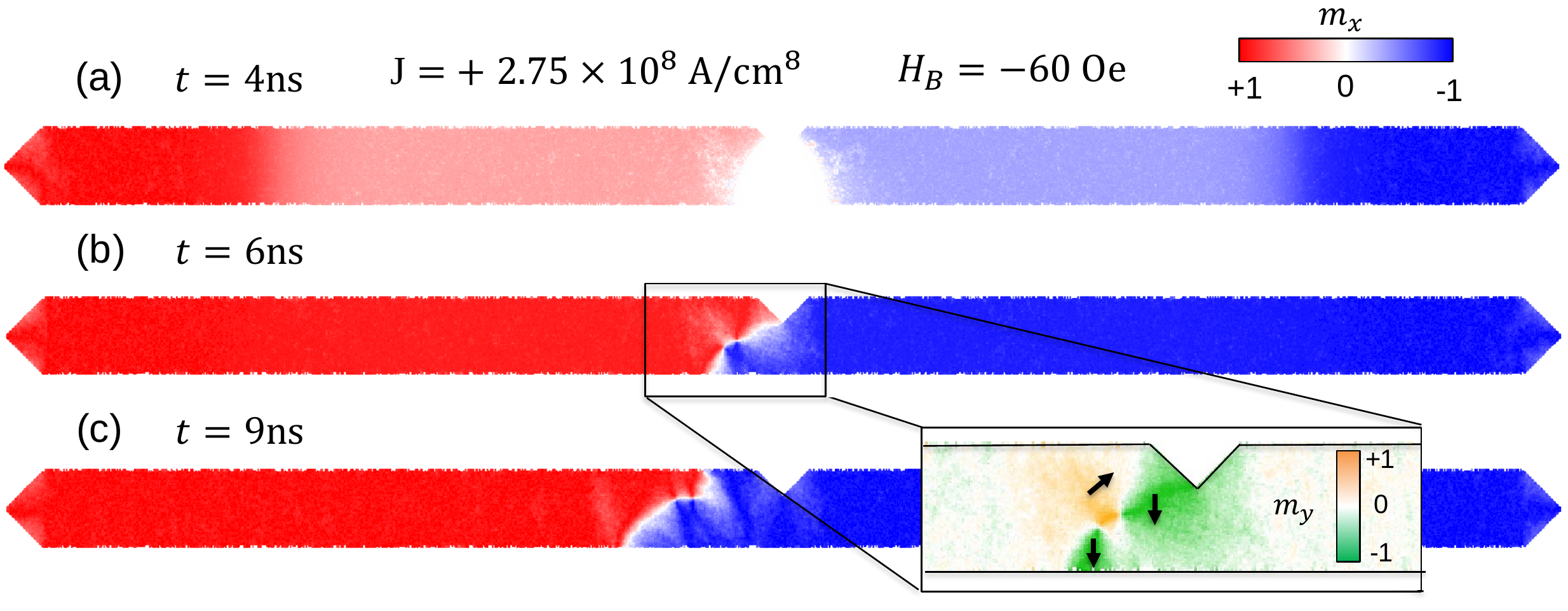}
\caption{(color on-line) Snapshots of the magnetization for $H_B=-60\ {\rm Oe}$ and $J=2.75\times10^8\ {\rm A/cm^2}$ at $t=4\mathrm{ ns}$, $6\mathrm{ ns}$ and $9\mathrm{ ns}$. \gg{The detail in (b) represents a meta-stable state formed after re-nucleation}.}
\label{fig:Fig5_MsnapT}
\end{figure}

\gg{Coming back to our system, }for $|J|=2.75\times10^8\ {\rm A/cm^2}$ (Fig.~\ref{fig:MoviesSnap}(b)), the average temperature between the contacts is $T_{av}\approx 740\ {\rm K}<T_C$ while the local temperature in the notch is larger than the Curie threshold ($T_{notch}\approx 980\ {\rm K}>T_C$). In this case, the destruction of the ferromagnetic order occurs only around the notch position, as shown in Fig.~\ref{fig:Fig5_MsnapT}(a) at $t=4\mathrm{ nm}$ and in the corresponding movie (Fig.~\ref{fig:MoviesSnap}(b)). After the pulse, DW is re-nucleated under the notch but adopts a different internal structure, \gg{the new structure can either be a displaced transverse (or vortex) wall or a meta-stable state which then collapses into a transverse or vortex wall.} An example of this after-effect configuration is shown in the transient magnetization states of Fig.~\ref{fig:Fig5_MsnapT} at $t=6\mathrm{ ns}$ and $t=9\mathrm{ ns}$. \gg{Note the meta-stable state shown in Fig.~\ref{fig:Fig5_MsnapT}b}. \b{The new DW is found to have a lower depinning field and it is eventually depinned from the notch. We found a new critical depinning field for the re-nucleated DW of about $45\ {\rm Oe}$. This decrease in the depinning field can be attributed to the new DW structure \gg{and the complex re-nucleation process} (note that the re-nucleation takes place under the effect of the bias field \gg{and can involve transitions from meta-stable states}).}  

Since these results clearly indicate a remarkable influence of Joule heating in current-assisted DW depinning, it is interesting to evaluate the magnetization dynamics coupled to the heat transport in the absence of STTs ($P^0=0$). This case is depicted by green diamonds in Fig.~\ref{fig:Fig4_PhaseD_LLB}. \gg{These latter results are very close to those observed in the presence of STTs (blue circles and red squares in Fig.~\ref{fig:Fig4_PhaseD_LLB}). Consequently, we can state that STTs do not play a dominant role for $|J|>2.5\times10^8\ {\rm A/cm^2}$. By contrast, for such high currents, Joule heating and the applied field are the main agents responsible for the DW depinning events.} For lower currents ($|J|<2.5\times10^8\ {\rm A/cm^2}$), where DW depinning is mainly achieved because of the applied field, the contribution of the STTs is also small, as can be deduced by considering the similarity of the results with and without STTs in Fig.~\ref{fig:Fig4_PhaseD_LLB}.

\section{Discussion}
\label{sec:discussion}
\b{As previously commented, temperature is coupled to the magnetization dynamics through the thermal fields and the temperature dependence of the magnetic parameters. However, by simulating the DW depinning without thermal fields (Eq.~\ref{eq:thfield})  we found that thermal fluctuations effect is negligible. Indeed we obtained almost the same depinning diagram even without the random thermal fields. The main actor is therefore the temperature dependence of the micromagnetic parameters, which is indeed responsible for the destruction of the ferromagnetic order and the previously described depinning events at high currents. However, even at $T<T_C$, temperature can significantly affect the depinning field. By performing purely field driven depinning simulations at constant uniform temperature, we found that the depinning field changes with $T$ as $H_d(T)\approx H_d^0 m_e(T)$, where $H_d^0$ is the depinning field at $T=0$. This means that temperature is actually decreasing the pinning barrier. \gg{Such behaviour is due to exchange and magnetostatic energies which scale as $m_e(T)^2$ and give rise to the pinning barrier. Since the Zeeman energy scales as $m_e(T)$, the resulting depinning field (defined as the field at which the Zeeman energy is equal to the energy barrier) scales as $m_e(T)$. Note that this behaviour is true for a system at constant and uniform temperature but, in our case,  time- and space-dependent temperature patterns can lead to further pinning and the behaviour of the depinning field might be more complicated. The analysis performed at constant uniform temperature simply help us to understand one of the possible effect of the temperature rise. Experimental observations (in uniform Py strips) of decreases in the DW propagation field~\cite{Yamaguchi2006} or in the switching field~\cite{Togawa2008} due to Joule heating, can be explained by the same mechanism, as already anticipated in Ref.~\cite{Togawa2008,Yamaguchi2006}} Note also that even if temperature effects can be divided into these two contributions (thermal noise and temperature dependence of the micromagnetic parameters) at a micromagnetic level, from a more fundamental and microscopic point of view, all of them are related to thermal agitation of single spins. 
Finally, although it is not the main objective of the present study, it is interesting to comment about one of the main results of Hayashi et al.~\cite{Hayashi2006a} which is the independence of the critical depinning current on the DW type (transverse or vortex) at low fields (high currents). Our analysis, although performed only one DW type, points out, in agreement with the conclusion of Ref.~\cite{Hayashi2006a}, that such independence might be due to thermal effects which overcome the STTs and are indeed independent on the DW type. 
}

\section{Conclusions}
\label{sec:conclusions}

The current-assisted DW depinning from a notch in a Permalloy strip on top of a Silicon substrate has been evaluated under static fields and short current pulses. The non-uniform spatial distribution of the current due to the notch results in significant non-uniform temperature profiles through the sample. Owing to Joule heating, the temperature can reach values close to or even above the Curie point for commonly injected currents in typical experiments. Although significant Joule heating have been observed in experimental studies, its effect has been overlooked in the theoretical descriptions. 
Here, we have developed a formalism to properly describe the magnetization dynamics coupled to the heat transport. In particular, the depinning diagrams of the critical depinning current as a function of the applied field were experimentally found to be insensitive to the current polarity in the system under study, an observation which is not compatible with the sole contribution of the spin transfer torques. Our analysis demonstrates that indeed Joule heating is crucial to reproduce these experimental observations~\cite{Hayashi2006a}. In agreement with previous studies~\cite{Fangohr2011,Ramos2015}, the temperature evolution of the strip strongly depends on the current amplitude and the substrate. Below the notch, temperature is much higher than the average value. The rise in temperature leads to an increase in thermal agitation and in a reduction of the depinning field. Moreover, under typically injected current pulses, Joule heating can lead to a local destruction of the ferromagnetic order during which the DW is destroyed and the re-nucleated with a different internal structure \gg{(vortex, transverse or a meta-stable state)} and a lower depinning field. 
Our findings suggest that previous estimations of STT parameters based on depinning experiments, performed by fitting the experimental data with conventional micromagnetic and/or 1D models at constant and uniform temperature, must be carefully considered in systems where Joule heating is relevant. In addition, the formalism introduced here can be used to study the interplay between STT and thermal gradients in systems where the temperature changes dynamically.

\begin{acknowledgments}
This work was supported by project WALL, FP7-PEOPLE-2013-ITN 608031 from the European Commission, projects MAT2011-28532-C03-01 and MAT2014-52477-C5-4-P from the Spanish government and projects SA163A12 and SA282U14 from the Junta de Castilla y Leon.
\end{acknowledgments}


\begin{thebibliography}{}
\bibitem{note5} \gg{An alternative could be importing the temperature distribution obtained by COMSOL at each time-step. However, also this process would significantly slow down the simulations since the copy of data from and to the GPU's memory requires a considerable time and it would need to be repeated many times ($\approx 10^5$ times). Considering the remarkable agreement with the phenomenological model, this operation is not needed.}
\bibitem{note4} \gg{The effect of $H_B$ has been neglected in the calculation since its contribution is negligible for the magnitude of the maximum applied field ($H_B^{max}\approx10$mT).}
\end{thebibliography}

\newpage

\end{document}